\documentclass[twocolumn,aps,amssymb,showpacs,superscriptaddress,longbibliography,10pt]{revtex4-1}
\usepackage{graphicx}
\usepackage{epsfig}
\usepackage{natbib}
\usepackage{dcolumn}
\usepackage{bm}
\usepackage{soul}
\usepackage{sidecap}
\usepackage[usenames,dvipsnames]{color}
\usepackage{setspace}
\usepackage[colorlinks=true,linkcolor=blue,urlcolor=blue,citecolor=blue]{hyperref}
 \usepackage[usenames,dvipsnames,svgnames,table]{xcolor}

\begin{document}
 
\newcommand{\lsco}{La$_{2-x}$Sr$_x$CuO$_4$}
\newcommand{\lbco}{La$_{2-x}$Ba$_x$CuO$_4$}
\newcommand{\lnsco}{La$_{1.6-x}$Nd$_{0.4}$Sr$_x$CuO$_4$}
\newcommand{\lesco}{La$_{1.8-x}$Eu$_{0.2}$Sr$_x$CuO$_4$}
\newcommand{\lresco}{La$_{2-x-y}$RE$_{y}$Sr$_x$CuO$_4$}

\newcommand{\lnscoxy}{La$_{2-x-y}$Nd$_{y}$Sr$_x$CuO$_4$}
\newcommand{\lnco}{La$_{1.6}$Nd$_{0.4}$CuO$_4$}
\newcommand{\pcco}{Pr$_{2-x}$Ce$_{x}$CuO$_{4}$}
\newcommand{\ybco}{YBa$_{2}$Cu$_{3}$O$_{y}$}
\newcommand{\tltwotwoone}{Tl$_{2}$Ba$_{2}$CuO$_{6+\delta}$}
\newcommand{\hbco}{HgBa$_{2}$CuO$_{4+\delta}$}
\newcommand{\bsco}{Bi$_2$Sr$_{2}$CuO$_{6+\delta}$}
\newcommand{\bscco}{Bi$_2$Sr$_{2}$CaCu$_{2}$O$_{8+\delta}$}

\newcommand{\muohmcm}[1]{#1\,$\mu\Omega$\,cm}

\newcommand{\TN}{$T_{\rm N}$}
\newcommand{\Tc}{$T_{\rm c}$}
\newcommand{\Tstar}{$T^{\star}$}
\newcommand{\TCDW}{$T_{\rm CDW}$}
\newcommand{\Hc}{$H_{\rm c2}$}
\newcommand{\HS}{$H_{\rm Seebeck}$}

\newcommand{\RH}{$R_{\rm H}$}
\newcommand{\Sa}{$S_{\rm a}$}
\newcommand{\Sb}{$S_{\rm b}$}

\newcommand{\ie}{{\it i.e.}}
\newcommand{\eg}{{\it e.g.}}
\newcommand{\etal}{{\it et al.}}


\title{Anisotropy of the Seebeck Coefficient 
in the Cuprate Superconductor YBa$_{2}$Cu$_{3}$O$_{y}$: Fermi-Surface Reconstruction by Bidirectional Charge Order}


\author{O.~Cyr-Choini\`{e}re}
\altaffiliation{Present address: Department of Physics, McGill University, Montreal, Qu\'{e}bec H3A 2T8, Canada}
\affiliation{D\'{e}partement de physique  \&  RQMP, Universit\'{e} de Sherbrooke, Sherbrooke,  Qu\'{e}bec J1K 2R1, Canada}

\author{S.~Badoux}
\affiliation{D\'{e}partement de physique  \&  RQMP, Universit\'{e} de Sherbrooke, Sherbrooke,  Qu\'{e}bec J1K 2R1, Canada}

\author{G.~Grissonnanche}
\affiliation{D\'{e}partement de physique  \&  RQMP, Universit\'{e} de Sherbrooke, Sherbrooke,  Qu\'{e}bec J1K 2R1, Canada}

\author{B.~Michon}
\affiliation{D\'{e}partement de physique  \&  RQMP, Universit\'{e} de Sherbrooke, Sherbrooke,  Qu\'{e}bec J1K 2R1, Canada}

\author{S.~A.~A.~Afshar}
\affiliation{D\'{e}partement de physique  \&  RQMP, Universit\'{e} de Sherbrooke, Sherbrooke,  Qu\'{e}bec J1K 2R1, Canada}

\author{S.~Fortier}
\affiliation{D\'{e}partement de physique  \&  RQMP, Universit\'{e} de Sherbrooke, Sherbrooke,  Qu\'{e}bec J1K 2R1, Canada}

\author{D.~LeBoeuf}
\affiliation{Laboratoire National des Champs Magn\'{e}tiques Intenses, UPR 3228, (CNRS-INSA-UJF-UPS), Grenoble 38042, France}

\author{D.~Graf}
\affiliation{National High Magnetic Field Laboratory, Tallahassee, FL 32310, USA}

\author{J.~Day}
\affiliation{Department of Physics and Astronomy, University of British Columbia, Vancouver, British Columbia V6T 1Z4, Canada}

\author{D.~A.~Bonn}
\affiliation{Department of Physics and Astronomy, University of British Columbia, Vancouver, British Columbia V6T 1Z4, Canada}
\affiliation{Canadian Institute for Advanced Research, Toronto, Ontario M5G 1Z8, Canada}

\author{W.~N.~Hardy}
\affiliation{Department of Physics and Astronomy, University of British Columbia, Vancouver, British Columbia V6T 1Z4, Canada}
\affiliation{Canadian Institute for Advanced Research, Toronto, Ontario M5G 1Z8, Canada}

\author{R.~Liang}
\affiliation{Department of Physics and Astronomy, University of British Columbia, Vancouver, British Columbia V6T 1Z4, Canada}
\affiliation{Canadian Institute for Advanced Research, Toronto, Ontario M5G 1Z8, Canada}

\author{N.~Doiron-Leyraud}
\affiliation{D\'{e}partement de physique  \&  RQMP, Universit\'{e} de Sherbrooke, Sherbrooke,  Qu\'{e}bec J1K 2R1, Canada}

\author{Louis~Taillefer}
\email{louis.taillefer@usherbrooke.ca}
\affiliation{D\'{e}partement de physique  \&  RQMP, Universit\'{e} de Sherbrooke, Sherbrooke,  Qu\'{e}bec J1K 2R1, Canada}
\affiliation{Canadian Institute for Advanced Research, Toronto, Ontario M5G 1Z8, Canada}

\date{\today}


\begin{abstract}

The Seebeck coefficient $S$ of the cuprate YBa$_{2}$Cu$_{3}$O$_{y}$ was measured in magnetic fields large enough to suppress superconductivity, 
at hole dopings $p$\,=\,$0.11$ and $p$\,=\,$0.12$, for heat currents along the $a$ and $b$ directions of the orthorhombic crystal structure. 
For both directions, $S$\,/\,$T$ decreases and becomes negative at low temperature, 
a signature that the Fermi surface undergoes a reconstruction due to broken translational symmetry. 
Above a clear threshold field, a strong new feature appears in $S_{\rm b}$, for conduction along the $b$ axis only. 
We attribute this feature to the onset of 3D-coherent unidirectional charge-density-wave modulations 
seen by x-ray diffraction, also along the $b$ axis only. 
Because these modulations have a sharp onset temperature
well below the temperature where $S$\,/\,$T$ starts to drop towards negative values,
%
we infer that they
are not the cause of Fermi-surface reconstruction.
%
Instead, the reconstruction must be caused by the quasi-2D bidirectional modulations that develop at significantly higher temperature.

\end{abstract}




\maketitle



\section{Introduction}
\label{sec:Intro}

In the last decade, various transport measurements in high magnetic fields
have revealed that the 
Fermi surface of hole-doped cuprate superconductors 
undergoes a reconstruction at low temperature in a doping interval centered at $p$\,$\simeq$\,$0.12$~\cite{Taillefer2009}.
The key feature of this Fermi-surface reconstruction (FSR) is the presence of a small electron-like pocket,
detected by quantum oscillations~\cite{Doiron-Leyraud2007,Yelland2008,Bangura2008,Barisic2013}, 
combined with sign changes in the temperature dependence of the Hall ($R_{\rm H}$)
and Seebeck ($S$) coefficients, from positive at high temperature to negative at low temperature. 
A negative \RH{} or $S$ has now been observed in seven hole-doped cuprates: 
\ybco{} (YBCO)~\cite{LeBoeuf2007,LeBoeuf2011,Chang2010,Laliberte2011},
YBa$_{2}$Cu$_{4}$O$_{8}$ (Y124)~\cite{LeBoeuf2007},
\hbco{} (Hg1201)~\cite{Doiron-Leyraud2013},
\lsco{} (LSCO)~\cite{Suzuki2002,Badoux2016a},
\lesco{} (Eu-LSCO)~\cite{Laliberte2011},
\lnsco{} (Nd-LSCO) \cite{Noda1999,Hucker1998},
and 
\lbco{} (LBCO) \cite{Adachi2001}.

There is compelling evidence that this FSR is caused by charge-density-wave (CDW) order.
Indeed, in all materials and at every doping where FSR has been detected,
CDW modulations have also been observed by x-ray diffraction (XRD)~\cite{Tranquada1995,Fink2011,Ghiringhelli2012,Chang2012a,Tabis2014,Croft2014}
(except in Y124, where no XRD search has been reported).
Having said this, the mechanism by which CDW order produces a small electron pocket in the Fermi surface
of hole-doped cuprates remains a puzzle.
This is because CDW order is thought to be unidirectional (or ``stripe-like") in at least some cuprates 
and a unidirectional CDW modulation does not in general produce a closed electron pocket~\cite{Millis2007},
at least not at ``nodal" locations in the Brillouin zone, away from the anti-nodal pseudogap~\cite{Yao2011}.
By contrast, bidirectional CDW order (with in-plane modulations along both high-symmetry directions
of the tetragonal or orthorhombic lattice) readily produces a closed electron pocket at ``nodal" locations~\cite{Harrison2012,Allais2014}.

This paradox has recently become vivid in the orthorhombic cuprate YBCO at $p$\,=\,$0.12$, 
where XRD studies in high magnetic fields detect long-range three dimensional (3D) CDW order~\cite{Gerber2015},
with modulations that run only along the $b$ axis~\cite{Chang2016,Jang2016}, above a sharply defined threshold field that coincides
with an anomaly in the sound velocity considered to be the thermodynamic signature of CDW order in YBCO~\cite{LeBoeuf2013}. 
%
Is this field-induced unidirectional CDW order causing the FSR in YBCO?

Here, we report measurements of the Seebeck coefficient $S$ of YBCO along the $a$ and $b$ axes at $p$\,=\,$0.11$ and $p$\,=\,$0.12$,
in magnetic fields high enough to reach the normal state. 
For both directions, we observe a negative $S$ at low temperature, signature of the FSR giving an electron pocket. 
In addition, we detect a pronounced minimum in $S_{\rm b}(H)$, not present in $S_{\rm a}(H)$,
whose onset field 
and temperature 
match the 
onset of the 3D unidirectional CDW order seen by XRD.
%
However, since this onset temperature is well below the temperature where $S$\,/\,$T$ starts to drop towards negative values,
we infer that the primary cause of the FSR are the 2D bidirectional CDW modulations that develop in tandem
with the gradual drop in $S$\,/\,$T$.
%



\begin{figure}
\centering
\includegraphics[width=0.46\textwidth]{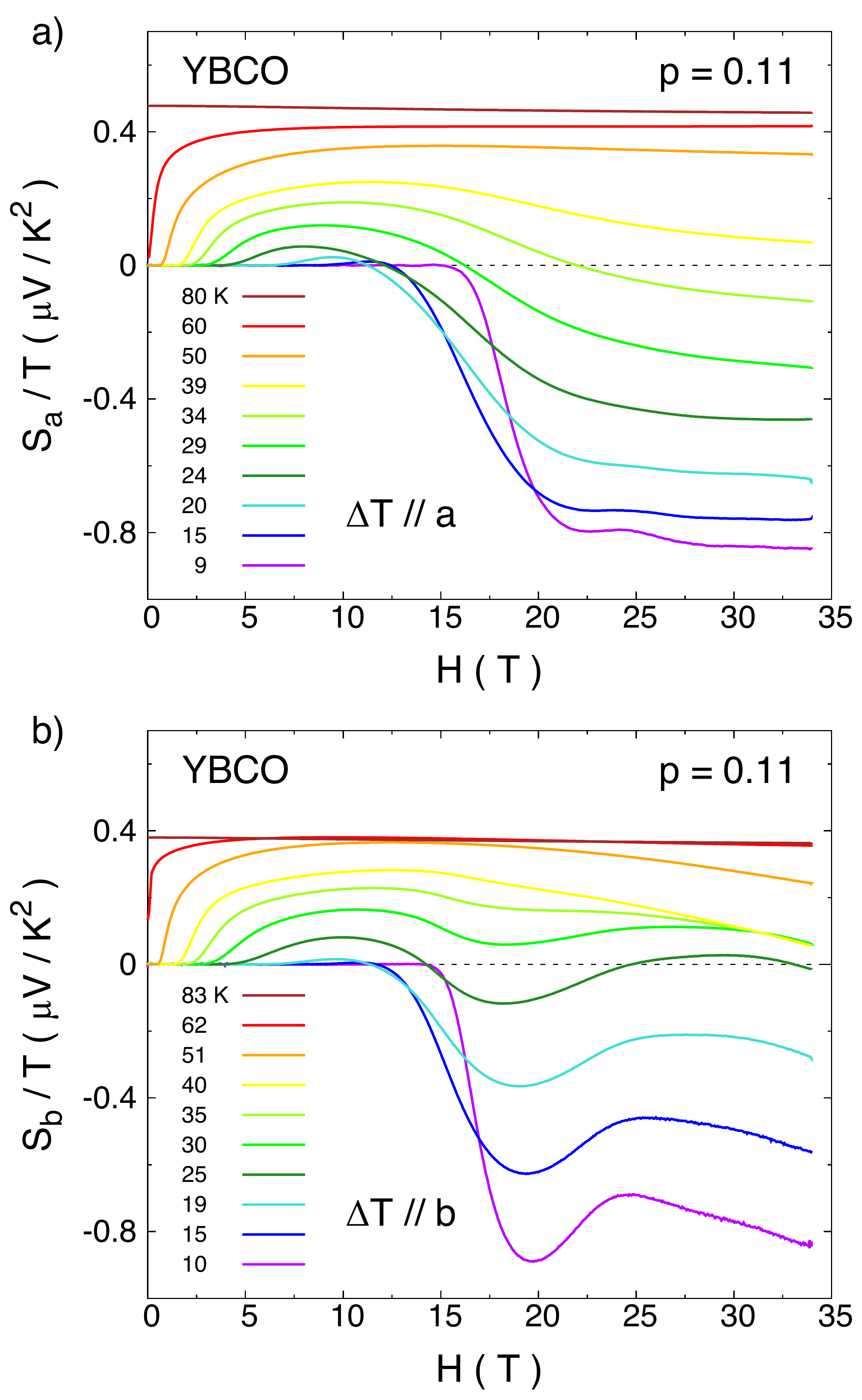}
\caption{Seebeck coefficient of YBCO at $p$\,=\,$0.11$, for a heat current along the $a$ axis (a) and $b$ axis (b) 
			  of the orthorhombic crystal structure,
			  plotted as $S$\,/\,$T$ vs magnetic field $H$, at various temperatures, as indicated.
			  The negative value of $S$\,/\,$T$ at low temperature and high field is the signature of Fermi-surface reconstruction.
			  In the isotherms of \Sb\,$/$\,$T$ vs $H$ (b), a clear dip develops below $T$\,$\simeq$\,$40$\,K, at $H$\,$\simeq$\,$18$\,T.
}
\label{SabvsHallT}
\end{figure}



\section{Methods}
\label{sec:Methods}

Single crystals of \ybco{} (YBCO) were prepared by flux growth~\cite{Liang2012}. 
Their hole concentration (doping) $p$ is determined from the superconducting transition temperature $T_{\rm c}$~\cite{Liang2006},
defined as the temperature below which the zero-field resistance is zero.  
A high degree of oxygen order was achieved for samples with $p$\,=\,$0.11$ ($y$\,=\,$6.54$, ortho-II order, \Tc\,=\,$61.5$\,K) and  
$p$\,=\,$0.12$ ($y$\,=\,$6.67$, ortho-VIII order, \Tc\,=\,$65.4$\,K). 
The Seebeck coefficient $S$ -- the longitudinal voltage generated by a longitudinal thermal gradient -- was measured, as described elsewhere~\cite{Laliberte2011}, 
on two pairs of $a$-axis and $b$-axis YBCO samples, with dopings $p$\,=\,$0.11$ and $p$\,=\,$0.12$. 
$S(H)$ was measured as a function of magnetic field up to $H$\,=\,$34$\,T,
applied along the $c$~axis,
in YBCO samples with $p$\,=\,$0.11$ at the 
Laboratoire National des Champs Magn\'{e}tiques Intenses, (LNCMI) in Grenoble 
and in samples with $p$\,=\,$0.12$ at the 
National High Magnetic Field Laboratory (NHMFL) in Tallahassee.
Our $b$-axis sample with $p$\,=\,$0.11$ was also measured up to $H$\,=\,$45$\,T, at the NHMFL. 
At $p$\,=\,$0.11$ and $0.12$, the critical field for suppressing superconductivity in YBCO is 
\Hc\,=\,$25$\,T~\cite{Grissonnanche2014}.



\begin{figure}
\centering
\includegraphics[width=0.46\textwidth]{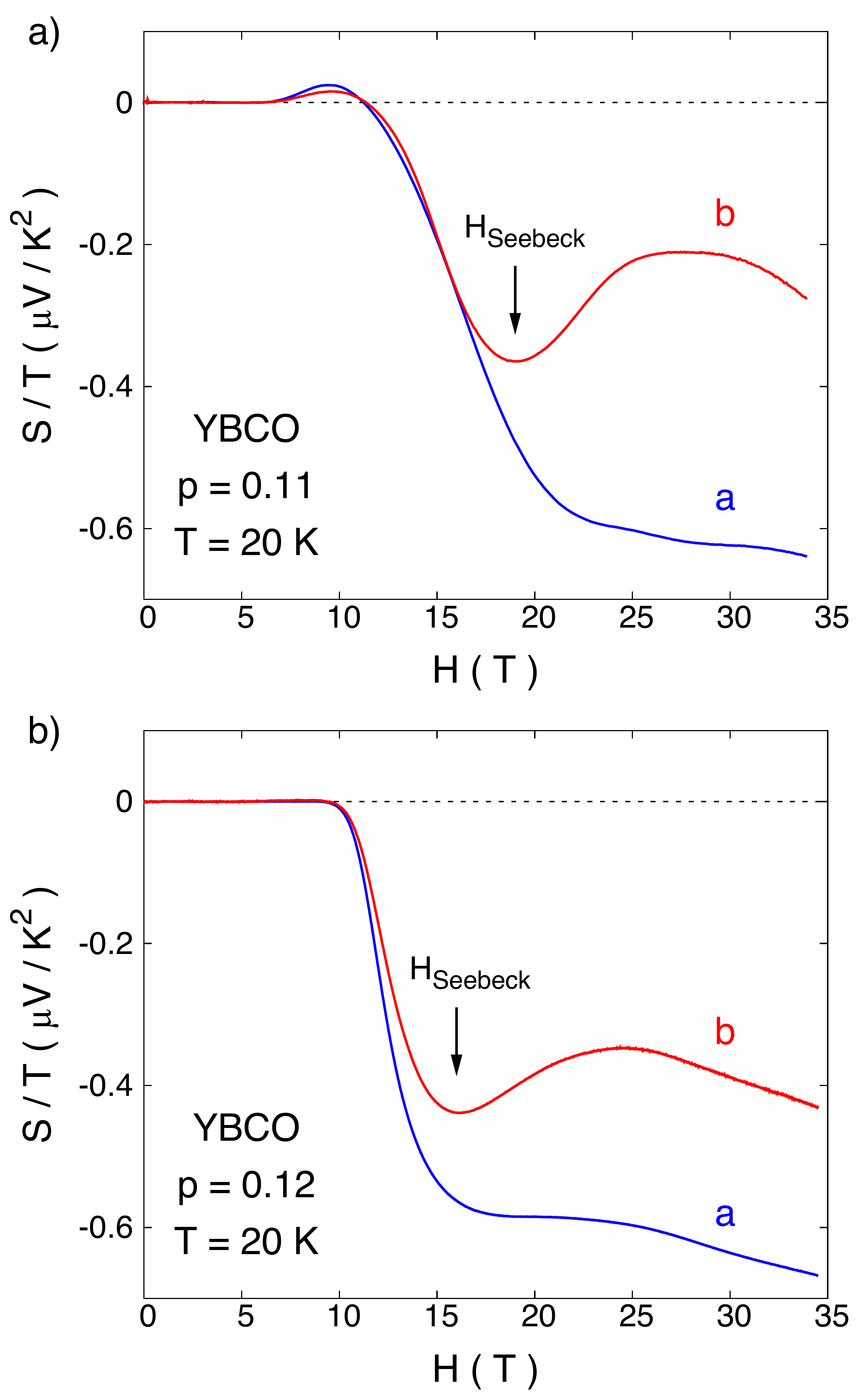}
\caption{Seebeck coefficient of YBCO along the $a$ axis (blue) and $b$ axis (blue) at $T$\,=\,$20$\,K,
			  for $p$\,=\,$0.11$ (a) and $p$\,=\,$0.12$ (b), plotted as $S$\,/\,$T$ vs magnetic field $H$. 
			  At both dopings, the Seebeck coefficient exhibits a strong anisotropy, manifest as an upturn 
			  in \Sb{} above $H_{\rm Seebeck}$ (arrow).
}
\label{SabvsH20K}
\end{figure}



\section{Results}
\label{sec:Results}

In Fig.~\ref{SabvsHallT}, the Seebeck coefficient of YBCO at $p$\,=\,$0.11$ 
is plotted as $S$\,/\,$T$ vs $H$, for several temperatures. 
Our data on \Sa{} agree well with previous measurements of the 
Seebeck coefficient in YBCO~\cite{Chang2010,Laliberte2011}. 
To our knowledge, there are no prior high-field measurements  of \Sb{} in YBCO. 
We see that for both directions, $S$\,/\,$T$ at high field goes from positive at high temperature
to negative at low temperature,
the signature that FSR is occurring upon cooling,
resulting in a Fermi surface at low temperature that contains a small electron pocket~\cite{Taillefer2009}. 
Note that the magnitude of $S$\,/\,$T$ at $T$\,$\rightarrow$\,$0$ ($\simeq$\,$-0.8$\,$\mu$V\,/\,K$^2$) 
is consistent with theoretical expectation~\cite{Behnia2004} in the sense that 
%
$S$\,/\,$T$\,=\,$-(\pi^2/2)$\,$(k_{\rm B}/e)$\,$( 1/T_{\rm F})$\,=\,$-1.0$\,$\mu$V\,/\,K$^{2}$~\cite{Laliberte2011},  
%
if we use the Fermi temperature $T_{\rm F}$\,=\,$410$\,K measured by quantum oscillations in YBCO at $p$\,=\,$0.11$~\cite{Doiron-Leyraud2007,Jaudet2008}.


\begin{figure}
\centering
\includegraphics[width=0.46\textwidth]{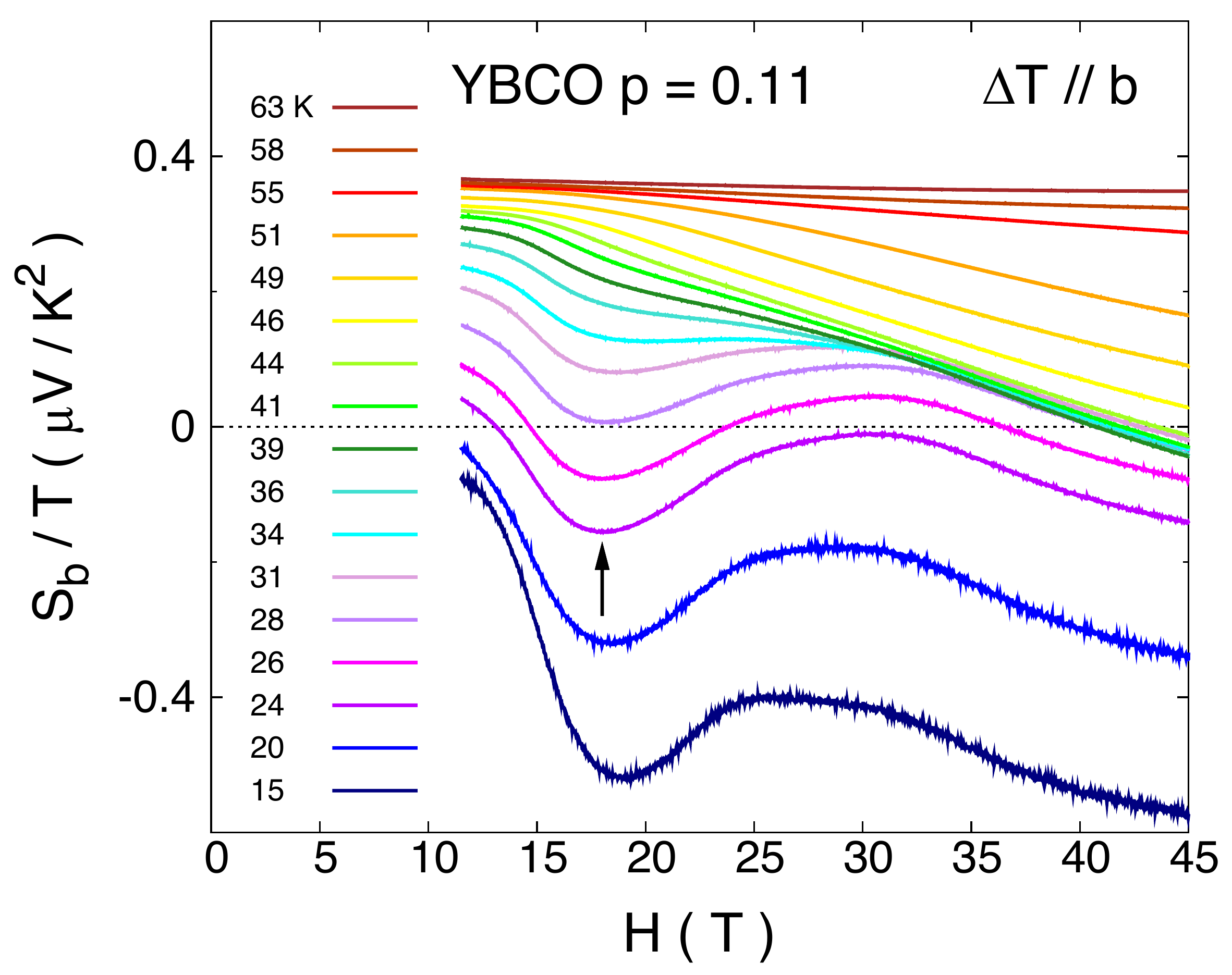}
\caption{Seebeck coefficient of YBCO at $p$\,=\,$0.11$ for a heat current along the $b$ axis, 
			  plotted as $S$\,/\,$T$ vs magnetic field $H$, at various temperatures, as indicated. 
			  This more complete data set complements that of Fig.~\ref{SabvsHallT}(b),
			  showing closely-spaced isotherms up to higher field.
			  The arrow marks $H_{\rm Seebeck}$, whose value is plotted in the $H-T$ phase diagram of Fig.~\ref{Phasediagram}.
			  A cut at $H$\,=\,$34$\,T yields the values of \Sb\,/\,$T$ plotted in Fig.~\ref{SbmfitvsT}(a)~(as red dots).
}
\label{SbvsHallT-TLH}
\end{figure}



\begin{figure}[t]
\centering
\includegraphics[width=0.46\textwidth]{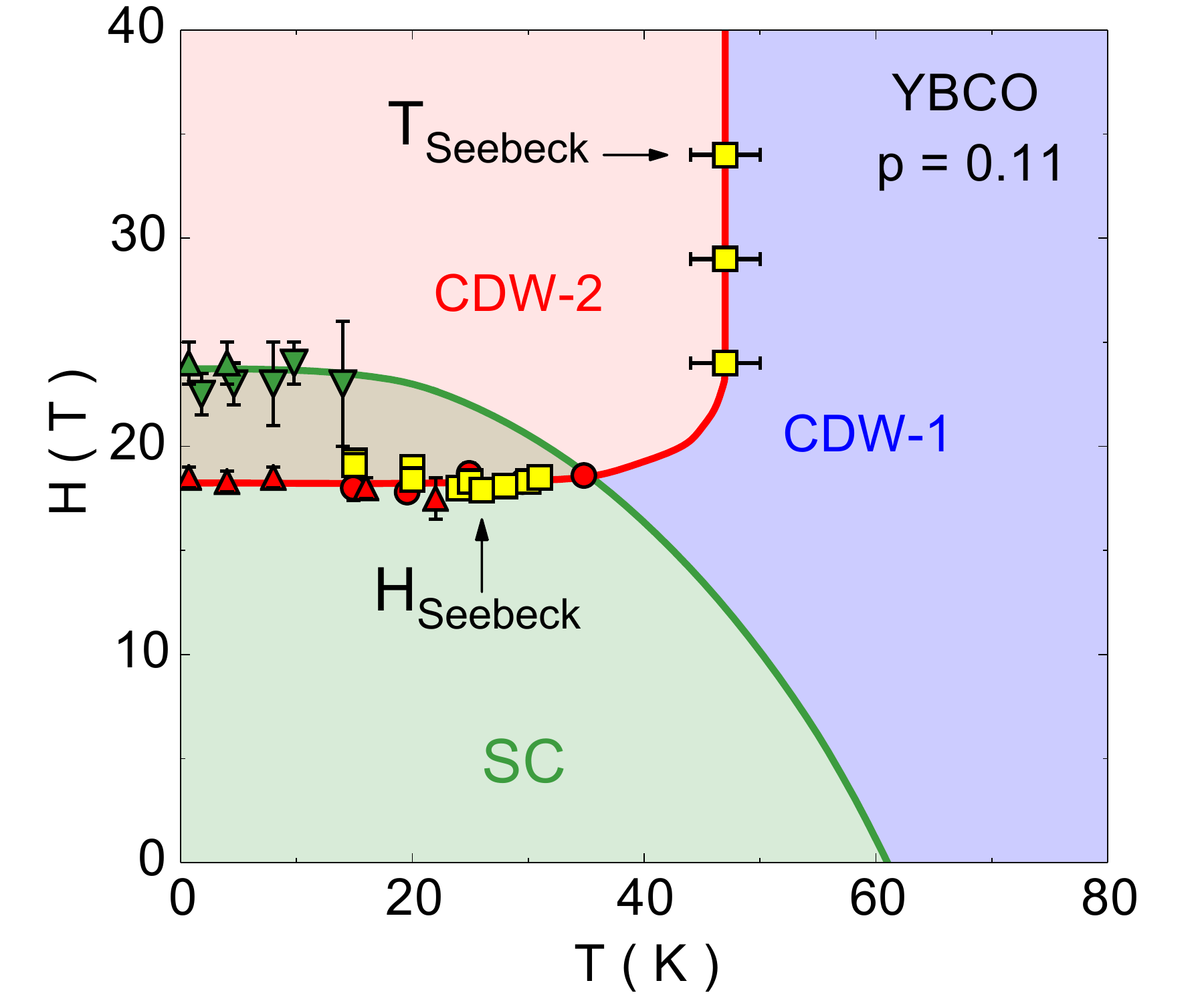}
\caption{Magnetic field-temperature phase diagram of YBCO at $ p$\,=\,$0.11$,
			  showing the field $H_{\rm Seebeck}$ (Fig.~\ref{SbvsHallT-TLH}) above which,
			  and the temperature $T_{\rm Seebeck}$ (Fig.~\ref{SbmfitvsT}) below which, the strong anomalous anisotropy in
			  the Seebeck coefficient appears (yellow squares).
			  $H_{\rm Seebeck}$ is seen to coincide with the threshold field
			  detected in the sound velocity (red circles; ref.~\onlinecite{LeBoeuf2013})
			  and in the thermal Hall conductivity $\kappa_{xy}$~(red triangles; ref.~\onlinecite{Grissonnanche2015}) of YBCO at $p$\,=\,$0.11$.
			  The upper critical field $H_{\rm c2}$ (green down-triangles from ref.~\onlinecite{Grissonnanche2014}; green up-triangles from ref.~\onlinecite{Grissonnanche2015}), 
			  is also shown, delineating the superconducting  (SC) phase.
			  The green line is a guide to the eye ending at the zero-field \Tc.
			  Short-range 2D bi-directional CDW modulations (CDW-1) are detected by XRD 
			  throughout this phase diagram, both in the blue region above the red and green lines, as well as below (to the left of) those lines.
			  In addition, long-range 3D unidirectional modulations (CDW-2) are detected in YBCO 
			  in a region very similar to the red region defined here, namely with an onset temperature
			 $T$\,$\simeq$\,$47$\,K and field $H$\,$\simeq$\,$18$\,T at $p=0.11$~\cite{Jang2016}, and
			 $T$\,$\simeq$\,$47$\,K and $H$\,$\simeq$\,$15$\,T at $p=0.12$~\cite{Chang2016}. 
			  The red line is a guide to the eye.
}
\label{Phasediagram}
\end{figure}



\begin{figure}[t]
\centering
\includegraphics[width=0.42\textwidth]{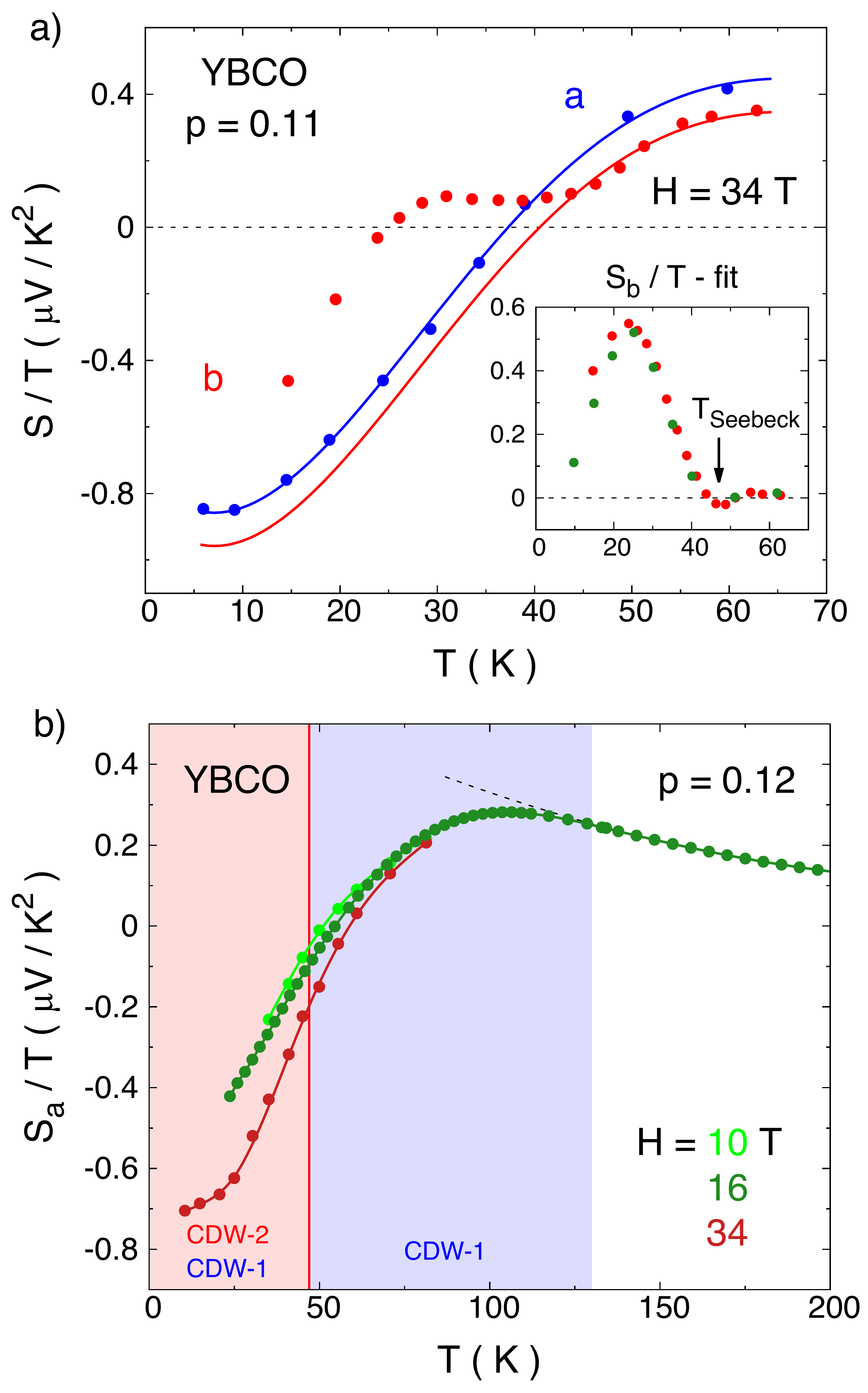}
\caption{a) Seebeck coefficient of YBCO at $p$\,=\,$0.11$ along the $a$ axis (blue) and $b$ axis (red),
			  at $H$\,=\,$34$\,T, plotted as $S$\,/\,$T$ vs $T$.
			  The blue line is a smooth fit to the \Sa{} data; 
			  the red line is the same line shifted  down by $0.1$\,$\mu$V\,/\,K$^2$. 
			  The anomalous feature seen in the field dependence of \Sb{} (Fig.~\ref{SabvsH20K}) 
			  shows up in the temperature dependence of \Sb\,/\,$T$ initially as a plateau.
			  \textit{Inset}: Difference between the $b$-axis data points and the red line (fit) in the top panel (red dots).
			  The green dots are obtained using the \Sb{} data of Fig.~\ref{SabvsHallT}(b).
			  The onset temperature for the extra anisotropy is $T_{\rm Seebeck}$\,=\,$47$\,$\pm$\,$5$\,K (arrow).
			  %
			  b) \Sa\,/\,$T$ vs $T$, in YBCO at $p$\,=\,$0.12$, for three field values as indicated.
			  At all fields, 2D~bi-directional CDW modulations (CDW-1) are observed in the blue and red regions~\cite{Chang2012},
			  while 3D~uni-directional CDW order (CDW-2) is observed only in the red region and
			  only when $H$\,$>$\,$15$\,T~\cite{Chang2016,Jang2016}.
			  The dashed line is a smooth extension of the high-$T$ data below its inflexion point at $T$\,=\,$130$\,K.
			  The dotted line is a linear extension of the data below $T$\,=\,$40$\,K.
}
\label{SbmfitvsT}
\end{figure}


The isotherms of \Sb{} in Fig.~\ref{SabvsHallT}(b) reveal a new and pronounced feature, essentially absent in \Sa. 
Indeed, on top of the same overall field and temperature dependence as observed in \Sa\,/\,$T$, 
\Sb\,/\,$T$ exhibits an upturn at high field, producing a dip at $H$\,$\simeq$\,$18$\,T
that deepens as temperature is reduced. 
In Fig.~\ref{SabvsH20K}(a), we focus on this feature by comparing \Sa\,/\,$T$ (blue) and \Sb\,/\,$T$ (red) vs $H$ at $T$\,=\,$20$\,K. 
At low field (up to about 16\,T), both curves are identical: 
zero in the vortex-solid state, 
then slightly positive, 
followed by a dramatic drop to large negative values. 
At fields above 16\,T, a striking anisotropy between the two directions appears,
as a pronounced upturn develops in \Sb, but not in \Sa. 
We identify the field at which \Sb{} reaches a minimum as $H_{\rm Seebeck}$, equal to $19$\,$\pm$\,$1$\,T at $T$\,=\,$20$\,K. 
Fig.~\ref{SabvsH20K}(b) presents the same comparison at $p$\,=\,$0.12$,
in crystals with a different oxygen order (ortho-VIII instead of ortho-II).
We observe a very similar Seebeck anisotropy, again characterized by an upturn in \Sb,
appearing above $H_{\rm Seebeck}$\,=\,$16$\,$\pm$\,$1$\,T.
 %

To study the temperature dependence of $H_{\rm Seebeck}$ in detail, we measured closely-spaced isotherms 
of \Sb{} up to $45$\,T, plotted in Fig.~\ref{SbvsHallT-TLH}.
We see that the minimum in \Sb\,/\,$T$ vs $H$ is present at temperatures up
to at least $30$\,K, remaining in roughly the same position.
In Fig.~\ref{Phasediagram}, we plot $H_{\rm Seebeck}$ on the $H-T$ phase diagram of YBCO at $p$\,=\,$0.11$ (yellow squares). 
It is essentially constant in temperature up to $30$\,K. 

In Fig.~\ref{SbmfitvsT}(a), \Sa$(T)$ and \Sb$(T)$ measured at $H$\,=\,$34$\,T are compared directly, 
plotted as $S$\,/\,$T$ vs $T$.
We see that down to $45$\,K, the two curves are approximately parallel, with a roughly constant difference between them.
Indeed, a smooth fit through the $a$-axis data (blue line) makes a good fit through the $b$-axis data
if the line is simply shifted down rigidly (red line).
Below $45$\,K, \Sa\,/\,$T$ continues its monotonic decrease,
but the anomalous feature in \Sb{} produces a striking departure of \Sb\,/\,$T$ from its fit line (red), 
initially as a plateau which persists down to $\simeq$\,$30$\,K.
To capture this extra anisotropy, we plot the difference between 
$b$-axis data and red fit line in the inset of Fig.~\ref{SbmfitvsT}(a). 
We see that it appears below $T_{\rm Seebeck}$\,=\,$47$\,$\pm$\,$5$\,K.
$T_{\rm Seebeck}$ is plotted on the $H-T$ phase diagram of Fig.~\ref{Phasediagram},
for three different fields.

\section{DISCUSSION}

The anomaly in \Sb{} we observe in YBCO at $p$\,=\,$0.11$ is confined to a region of the $H-T$ diagram (Fig.~\ref{Phasediagram}) 
that is essentially the same region where 3D unidirectional CDW order has been observed by XRD~\cite{Gerber2015,Chang2016,Jang2016}.
This order was detected in YBCO  above an onset field $H$\,=\,$18$\,T at $p$\,=\,$0.11$~\cite{Jang2016} 
and above $H$\,=\,$15$\,T at $p$\,=\,$0.12$~\cite{Gerber2015,Chang2016}, 
in good agreement with $H_{\rm Seebeck}$\,=\,$19$\,$\pm$\,$1$\,T and $16$\,$\pm$\,$1$\,T
at $p$\,=\,$0.11$ and $0.12$, respectively (Fig.~\ref{SabvsH20K}). 
This value of $H_{\rm Seebeck}$ at $p$\,=\,$0.11$ is 
in agreement with the anomaly in the sound velocity~\cite{LeBoeuf2013} that marks the phase transition to CDW order (Fig.~\ref{Phasediagram}).
Those field values are also in agreement with the threshold field detected in the thermal 
Hall conductivity $\kappa_{xy}$~\cite{Grissonnanche2015}, for both $p$\,=\,$0.11$ (Fig.~\ref{Phasediagram}) and $p$\,=\,$0.12$.
Therefore, it is clear that $H_{\rm Seebeck}$ coincides with the onset of 3D unidirectional CDW order.
 
The onset temperature for that order ($T$\,$\simeq$\,$47$\,K)~\cite{Chang2016,Jang2016}
is not far below the onset of the NMR splitting associated with CDW order~\cite{Wu2013}.
There is little doubt that $T$\,$\simeq$\,$47$\,$\pm$\,$5$\,K
coincides with the onset of 3D unidirectional CDW order.

The fact that we can clearly detect the onset of 3D unidirectional CDW order in the Seebeck coefficient 
allows us to examine whether it causes the FSR in YBCO.
In Fig.~\ref{SbmfitvsT}(b), we plot \Sa\,/\,$T$ vs $T$ at $p$\,=\,$0.12$ 
for $H$\,=\,$10$, $16$ and $34$\,T.
We see that \Sa$\,/\,T$ starts to deviate downward from its high-temperature behaviour below $T$\,$\simeq$\,$130$\,K,
it peaks at $105$\,K and then it drops to become negative below $\sim$\,$60$\,K.
This is a gradual process, which starts in parallel with the gradual growth of short-range 2D CDW modulations
seen in XRD below $\simeq$\,$140$\,K~\cite{Chang2012a}.
(Down to \Tc, both the Seebeck and the XRD intensity are independent of magnetic field.)

The decrease in $S$\,/\,$T$ upon cooling is the signature of the FSR that leads to the formation of a small electron 
pocket in the Fermi surface at low temperature, detected via quantum oscillations, whose Fermi energy is consistent 
with the value of $S$\,/\,$T$ at $T$\,$\to$\,$0$ (Fig.~\ref{SbmfitvsT}(b))~\cite{Laliberte2011}.
In other words, the entire evolution of \Sa$\,/\,T$ vs $T$ is quantitatively consistent (in temperature and in amplitude) with 
a scenario whereby FSR is caused by the 2D CDW modulations (CDW-1).
%
The fact that this evolution is completely unaffected by the sharp onset of the 3D unidirectional CDW order at 
$47$\,K, measured in YBCO at the same doping ($p$\,=\,$0.12$) and the same field ($H$\,=\,$16$\,T)~\cite{Chang2016},
indicates that it doesn't play a fundamental role in causing the FSR.
It appears to only confer an extra anisotropy.


The fact that the 2D CDW modulations are bi-directional, \ie~that they run along both the $a$ and $b$ directions in the CuO$_2$ planes
of YBCO, provides a natural mechanism for the formation of a small electron pocket in the reconstructed Fermi surface~\cite{Harrison2012,Allais2014},
located in nodal positions where the states are believed to be in underdoped cuprates with an anti-nodal pseudogap.
An analysis of the anomalies in the sound velocities concluded that the order responsible for the observed transition 
must be bi-directional~\cite{LeBoeuf2013}.

Note that the CDW modulations observed in Hg1201~\cite{Tabis2014} are very similar to the 2D CDW modulations in YBCO,
and they cause a very similar FSR~\cite{Chan2016}, with negative Hall and Seebeck coefficients at low temperature~\cite{Doiron-Leyraud2013}.
Therefore, attributing the cause of the FSR to these 2D CDW modulations is consistent 
with the fact that so far no field-induced 3D CDW order has been observed in Hg1201.

Given that 2D CDW modulations exist in the superconducting state at $H$\,=\,$0$ \cite{Ghiringhelli2012,Chang2012a},
one might ask: are there signatures of the FSR inside the superconducting phase,
\ie{} inside the green region of the $H-T$ phase diagram (Fig.~\ref{Phasediagram})?
The answer is yes: 
in YBCO at $p$\,=\,$0.11$, \RH{} at $T$\,=\,$15$\,K is negative for all fields down to 
$H$\,=\,$H_{\rm vs}$\,$\simeq$\,$10$\,T, the field below which the vortex solid forms and \RH\,$=0$~\cite{Grissonnanche2015}.
So a negative \RH{} is observed even when $H$\,$<$\,$H_{\rm Seebeck}$.
In the vortex-liquid state between $H_{\rm vs}$ and \Hc, the negative \RH{} could come from states inside the vortex core.

On the other hand, the thermal Hall conductivity $\kappa_{xy}$ is dominated by $d$-wave quasiparticles outside the vortex cores.
In YBCO at $p$\,=\,$0.12$, $\kappa_{xy}$ is negative in the normal state just above \Tc~\cite{Grissonnanche2015}, 
even in the limit $H$\,=\,$0$,
as is the electrical Hall conductivity \cite{LeBoeuf2007}.
Immediately below \Tc, $\kappa_{xy}$ becomes positive~\cite{Grissonnanche2015}.
This sudden change of sign could be due to a sudden increase in the quasiparticle mean free path as the inelastic scattering is gapped out,
as found in YBCO immediately below \Tc~\cite{Zhang2001}.
Because the correlation length 
of the 2D CDW modulations is rather short in YBCO (and even shorter in Hg1201),
the longer electronic mean free path in the superconducting state may well average over the short-range CDW and wipe out the FSR.
Increasing the field to suppress superconductivity makes $\kappa_{xy}$ negative again~\cite{Grissonnanche2015}.
The threshold field at which this change of sign happens coincides with $H_{\rm Seebeck}$ (Fig.~\ref{Phasediagram}),
\ie{} with the onset field for 3D CDW order.
This can be understood as follows:
3D CDW order competes with superconductivity, its onset precipitates the demise of superconductivity,
which causes a reduction in the mean free path, making the FSR by short-range 2D CDW modulations possible again.
In other words, 3D CDW order triggers the transition out of the superconducting phase and this is where
$\kappa_{xy}$ starts its transition from zero to its normal-state (negative) value~\cite{Grissonnanche2015}.

\section{SUMMARY}

In summary, the Seebeck coefficient $S$ of YBCO at $p$\,=\,$0.11$ and $0.12$ responds to two aspects of the 
complex CDW ordering  in this material.
First, as temperature is decreased from room temperature, \Sa\,/\,$T$ deviates gradually downward from its 
dependence at high temperature in parallel with the gradual growth in the 2D bi-directional CDW modulations 
detected by XRD well above \Tc.
\Sa\,/\,$T$ decreases below $T$\,$\simeq$\,$100$\,K to eventually become negative, extrapolating to a large
negative value at $T$\,$\to$\,$0$ that is quantitatively consistent with the small electron pocket in the normal-state 
Fermi surface detected by quantum oscillations at low temperature. 
We infer that the 2D bi-directional CDW modulations reconstruct the Fermi surface of YBCO, 
and produce the electron pocket.
The same is true for Hg1201.

Secondly, a pronounced anomaly appears in \Sb{} below a temperature and above a field
that are both consistent with the onset temperature and field of the 3D unidirectional CDW order
detected in YBCO 
by high-field XRD at $p$\,=\,$0.11$ and $0.12$.
We conclude that the extra anisotropy is due to that low-temperature order, which is not, however,
the primary cause of the FSR.
Nevertheless, given that the two types of CDW modulations (CDW-1 and CDW-2 in Fig.~\ref{Phasediagram}) 
have the same wavelength, they most likely have a common origin.
It would be helpful to further elucidate the nature of their interplay.



\begin{acknowledgments}

A portion of this work was performed at the Laboratoire National des Champs Magn\'{e}tiques Intenses of the CNRS, member of the European Magnetic Field Laboratory. 
Another portion of this work was performed at the National High Magnetic Field Laboratory, which is supported by 
the National Science Foundation Cooperative Agreement No. DMR-1157490, the State of Florida, and the U.S. Department of Energy. 
O.C.C. was supported by a fellowship from the Natural Sciences and Engineering Research Council of Canada (NSERC). 
D.L. thanks Agence Nationale de Recherche (UNESCOS project ANR-14-CE05-0007), 
the Laboratoire d'Excellence LANEF (ANR-10-LABX-51-01) and the Universit\'{e} Grenoble-Alpes (SMIng-AGIR) for their support.
L.T. thanks ESPCI-ParisTech, Universit\'{e} Paris-Sud, CEA-Saclay and the Coll\`{e}ge de France for their hospitality and support, 
and the \'{E}cole Polytechnique (ERC-319286 QMAC) and LABEX PALM (ANR-10-LABX-0039-PALM) for their support, while this article was written. 
R.L., D.A.B. and W.N.H. acknowledge funding from the Natural Sciences and Engineering Research Council of Canada (NSERC). 
L.T. acknowledges support from the Canadian Institute for Advanced Research (CIFAR) and funding from 
the Natural Sciences and Engineering Research Council of Canada (NSERC; PIN:123817), 
the Fonds de recherche du Qu\'{e}bec - Nature et Technologies (FRQNT), 
the Canada Foundation for Innovation (CFI),
and a Canada Research Chair. 
Part of this work was funded by the Gordon and Betty Moore Foundation's EPiQS Initiative (Grant GBMF5306 to L.T.).

\end{acknowledgments}



%


\end{document}